\documentstyle[epsfig,aps,prb,twocolumn,floats,eqsecnum,amssymb]{revtex}
\begin{document}

\bibliographystyle{prsty}

\title{
\begin{flushleft}
{\small PHYSICAL REVIEW B 
\hfill 
VOLUME {\normalsize 53,} NUMBER {\normalsize 17} 
\hfill
{\normalsize 1} MAY {\normalsize 1996}-I,
{\normalsize  11593--11605
%\renewcommand{\thefootnote}{\fnsymbol{footnote}}
%\footnotemark[2]
}
}\\
\end{flushleft}  
Self-consistent Gaussian approximation for classical spin systems: \\
Thermodynamics
}

\author{D.~A. Garanin}

\address{
I. Institut f\"ur Theoretische Physik, Universit\"at Hamburg,
Jungiusstr. 9, D-20355 Hamburg, Germany; \\
e-mail: garanin@physnet.uni-hamburg.de\\
%}
%\date{\today}
%\maketitle
%\abstract{
\smallskip
{\rm(Received 27 June 1995; revised manuscript received 26 December 1995)}
\bigskip\\
\parbox{14.2cm}
{\rm
\hspace{3ex} The self-consistent Gaussian approximation (SCGA) 
for classical spin systems 
described by a completely anisotropic $D$-component vector model
is proposed, which takes into account fluctuations of the 
molecular field and thus is a next step beyond the molecular field 
approximation. The SCGA is sensitive to the lattice 
dimension and structure and to the form of spin interactions 
and yields rather accurate values of the
field-dependent magnetization $m(H,T)$ and other thermodynamic functions 
in the whole plane $(H,T)$ excluding the vicinity of the critical point 
$(0,T_c)$, where the SCGA breaks down, showing a first-order phase transition. 
The values of $T_c$ themselves can be 
determined in the SCGA with an accuracy better than 1\% for actual
3-dimensional structures. At low and high temperatures the SCGA
recovers the leading terms of the spin-wave theory, the low- and
high-temperature series expansions, respectively. 
The accuracy of the SCGA increases with the
increase of the spin dimension $D$, and in the limit
$D\to\infty$ the exact solution for the spherical model is recovered.  
\medskip\\
%\pacs{PACS numbers: 05.50.+q, 75.10.-b, 75.10.Hk, 75.40.Cx}
} 
} 

\maketitle

%\renewcommand{\thefootnote}{\fnsymbol{footnote}}
%\renewcommand{\footnoterule}{\rule{0cm}{0cm}}
%\footnotetext[2]{Printed in Hamburg}

\section{INTRODUCTION}

The models considering spin as a classical vector variable of a fixed 
length are the most studied ones in the theory of temperature-induced 
phase transitions on a lattice. The reason for this is that quantum effects 
that are always present in real magnetic systems and can make 
calculations more intricate play, however, a secondary role 
at (and above) $T_c$ and do not change the critical behavior. Classical 
models are also a good approximation for magnetics with large spin values  
$S$, such as, e.g., the Heisenberg systems EuO and EuS having $S=7/2$. 
Generally, for systems 
described by the Heisenberg model, H($S$), with 
$S\gg 1$ quantum effects become irrelevant in the temperature range 
$T\gtrsim T_c/S$, where the whole Brillouin zone is populated by spin 
waves and their occupation numbers become large. Additionally, the 
arbitrary-$S$ Ising model, I($S$), without a transverse field 
can be treated with the same methods as classical ones, because 
no spin commutators appear and quantum effects are trivial. An 
advantage of classical vector models is that they can be formulated for 
an arbitrary number of spin components, 
as was done by Stanley \cite{sta68prl,sta68pr,sta74}. Such a 
generalization is important since some magnetics with a complicated 
structure possess an order parameter with $n>3$ symmetric 
components (see Refs.\onlinecite{mukkri75,antsok94}, and references therein).

In the absense of an analytical solution to the phase transition problem 
in three dimensions such numerical methods as 
low-temperature series expansions (LTSE's) 
for the I(1/2) model \cite{essfis63}  and 
high-temperature series expansions (HTSE's) 
for the I(1/2) \cite{domsyk57,bak61,fis63tab,gofetal93}, 
I($S$) \cite{sauworjas75,camdyk75prb},
H(S) \cite{ruswoo5863,rusbakwoo74}, 
H($\infty$) \cite{camdyk75jpc,adlholjan93} and 
classical plane rotator and  \mbox{$x$-$y$}  models \cite{fermoowor73},
as well as for the general $n$-component vector model, O($n$)
\cite{sta68prl,sta74,luewei88plb,luewei88npb,butcom95}, 
were successfully applied for an accurate calculation of thermodynamic 
quantities in a wide temperature range including the vicinity of $T_c$.
It gave the results for the critical indices of magnetic systems 
and favored the creation of scaling and universality concepts.
With the development of computational facilities and algorithms the 
series methods were permanently improved. 
As the latest benchmark the recent calculation \cite{butcom95}
of the HTSE series for the reduced susceptibility $T\chi(T)$ of the O($n$) 
models up to $(J/T)^{19}$ can be considered. 
Another very efficient numerical method competing with series 
expansions is based on Monte Carlo (MC) simulations
(see, e.g., Refs.\onlinecite{gothas93,gofetal93}). An extraction 
of accurate results for infinite systems from the simulation data
for the lattices with a finite linear dimension $L$
is based usually on the finite-size scaling \cite{fisbar72}. An 
alternative approach also using simulations is the chiral perturbation 
theory in powers of $1/L$ (see, e.g., Ref.\onlinecite{tomyon95} and
references therein). 

Lately the ideas of the statistical theory of magnetism together with 
the methods of calculation have penetrated into the field theory. 
In particular, the lattice-regularized scalar Higgs model 
in the chiral limit, which can be identified 
with the four-component classical Heisenberg model, O(4), in four dimensions
\cite{hasetal87,luewei88plb}, was 
studied with the HTSE \cite{luewei88plb,luewei88npb} and 
MC simulation \cite{hasetal87,goejanneu91} methods. 
Recently Wilczek and Rajagopal \cite{wil92,rajwil93} have related the 
two-flavor quantum chromodynamics (QCD) to the O(4) vector model in three 
dimensions, the gauge coupling constant $g^2$ determining the 
temperature and the quark mass $m_q$ being proportional to the applied 
magnetic field. This has initiated an extensive numerical work 
(see Refs.\onlinecite{karlae94,ber94,kankay95} and references therein).

Although HTSE's produce the series coefficients usually with the help of 
such diagram methods as the linked cluster expansion (LCE) 
\cite{eng63,wor74}, the results are represented as a sum of ``bare'' 
(unrenormalized) diagrams, each proportional to some power of $J/T$. 
Alternatively, there were attempts starting from the early years to sum 
up some ``important'' infinite diagrams series to obtain a closed-form 
equation for a magnetic system in terms of renormalized diagrams, which 
should be a good analytical approximation in the whole temperature range. 
It was shown, in particular, how the mean field approximation (MFA) can 
be obtained diagrammatically (see, e.g., Refs.\onlinecite{bro60,horcal61}). A 
further renormalization of diagrams for the Ising model by 
Horwitz and Callen \cite{horcal61} led to an improvement of the MFA  taking 
into account 
self-consistently Gaussian fluctuations of the molecular field. This 
important work remained seemingly unappreciated, 
since the resulting equations were not 
numerically investigated in a satisfactory way and the real accuracy of 
the approximation was not recognized. Only much later was this self-consistent 
Gaussian approximation (SCGA) for the Ising systems independently 
rediscovered and numerically analyzed in Ref.\onlinecite{garlut84i}. 

The methods developed by Horwitz and Callen \cite{horcal61} for the 
Ising model were generalized for the quantum Heisenberg model in the 
subsequent paper, Ref.\onlinecite{stietal63}. This first version of the spin 
diagram technique (SDT) had not, however, succeeded in 
formulating a Gaussian approximation for the Heisenberg systems, since 
due to quantum effects transverse spin cumulants acquire a time 
dependence and cannot be renormalized in a desirable way. 

Later a similar diagram technique was 
formulated independently in Refs.\onlinecite{vlp6767,izykasskr74} and 
further developed in Refs.\onlinecite{garlut84pa,garlutpan92}. 
Although SDT allows one to write down diagrammatic perturbation 
series for all temperatures in a regular way and recovers the known 
spin-wave and LTSE 
results in the ordered state, as well as the HTSE ones above $T_c$,
summation of nontrivial diagram sequences in all 
orders of a perturbation theory (apart from the usual cases of 
Dyson and vertex equations) seems to be impossible. Due to the use of 
the Wick theorem for the calculation of averages of transverse spin 
components the number and complexity of 
diagrams increase dramatically with each order and most of 
diagrams are divergent at $T>T_c$ and compensate each other only in 
final expressions. The latter is not the case only for Ising systems, 
where there are no problems with the noncommutativity of different spin 
components and quantum effects are trivial. In the classical limit
$S\to\infty$, which is of a primary importance in the theory of 
temperature-induced phase transitions, the quantum SDT does not 
essentially simplify.

In Ref.\onlinecite {garlut84d} an alternative diagram technique for classical 
spin 
systems was proposed, which explicitly takes advantage of their 
classical properties and is much simpler than the quantum SDT. 
It allows, in particular, calculation of 
thermodynamic quantities of a system without dealing with its dynamics. 
In the static case all spin components can be treated similarly, and the 
consideration can be carried out for a generalized completely 
anisotropic model of 
$D$-component classical spin vectors ($|{\bf m}|=1$) on a lattice:
\begin{equation}\label{dham}
{\cal H} = -{\bf H}\sum_{i}{\bf m}_i -
 \frac{1}{2}\sum_{ij}J_{ij}\sum_{\alpha=1}^D \eta_\alpha 
m_{\alpha i} m_{\alpha j} .
\end{equation}
If the exchange interaction $\eta_\alpha J_{ij}$ is isotropic, 
i.e., all anisotropy factors $\eta_\alpha=1$, this 
model reduces to the one pioneered by 
Stanley \cite{sta68prl,sta68pr,sta74}, 
who proved \cite{sta68pr} that it is in the limit $D\to \infty$ 
equivalent to the exactly soluble 
spherical model \cite {berkac52}. An important particular case 
of the general model (\ref{dham}) is the so-called \mbox{$n$-$D$} model \cite 
{garlut84d}, where $n\leqslant D$ spin components are coupled by the 
exchange interaction with an equal strength and the rest $D-n$ ones are 
``free'' (i.e., $\eta_\alpha=1$ for $\alpha \leqslant n$ and
$\eta_\alpha=0$ for $\alpha > n$). The \mbox{$n$-$D$} model contains as 
particular cases the $S=1/2$ Ising model, I(1/2), for $n=D=1$, the 
classical Ising model, I($\infty$), for $n=1$, $D=3$, the plane rotator 
model for $n=D=2$, the classical \mbox{$x$-$y$} model for $n=2$, $D=3$, and 
the classical Heisenberg model, H($\infty$), for $n=D=3$. The variable 
$n$ is the number of the order parameter components and determines the 
universality class of a system. The total number of spin components,
$D$, enters only such nonuniversal quantities as $T_c$. It is clear 
that the expansion of the critical indices for the large number of 
components can be only the $1/n$ expansion. To the contrary, we shall 
see below that the absolute values of thermodynamic quantities are 
naturally developed in powers of $1/D$ for $D\gg 1$, which is not 
automatically the same as $1/n$ for $n\neq D$.

In Ref.\onlinecite {garlut84d} the self-consistent Gaussian 
approximation by Horwitz and Callen was generalized for systems 
with continuous spin symmetry and 
it was shown that in the limit $D\to\infty$ the SCGA becomes 
exact and yields the solution of the spherical model, whereas all other 
diagrams die out as at least $1/D$. Accordingly, the SCGA becomes more 
accurate for high spin dimensions $D$ and works better for H($\infty$) 
model ($n=D=3$) than for the I(1/2) one ($n=D=1$).
Numerical calculations for I($S$) \cite{garlut84i}
and H($\infty$) \cite{garlut86jpf} models have shown 
that for different 3-dimensional lattice structures the SCGA yields the 
magnetization $m$ and other thermodynamic quantities in the whole 
temperature range excluding the close vicinity of $T_c$ 
with an overall accuracy about 1\%, 
including the determination of $T_c$ itself. 

In Refs.\onlinecite{garlut84i,garlut84d,garlut86jpf}, the SCGA was only
briefly described, and its analytical properties need to be
explained in more detail. Principally important is to test
the SCGA on models with lattice dimensionality $d\geqslant 4$
(hypecubic lattices) and to compare its results with those of
the $1/d$ expansion \cite{fisgau64,gerfis74}
and MC calculations \cite{gofetal93}.
In this case the SCGA should be more accurate,
since nontrivial effects of the fluctuation interaction
(i.e., non-Gaussian effects) die out \cite{larkhm69}.
In view of applications in the field theory mentioned above it is 
important to extend calculations to O($n$) models ($n=D$) with 
$n\geqslant 4$ and to make a comparison with the $1/n$ expansion
\cite{abe73,abehik7377,okamas78}. Some other
tasks are to perform a numerical solution of 
the SCGA equations in the case of a nonzero magnetic field, 
to make a comparison with the experimental data on
Eu chalcogenides, and to consider the lattices with the
next nearest neighbor (nnn) interactions.    
The solution of the problems mentioned above, 
as well as a detailed statement of the SCGA, is the aim of the 
present article.

In Sec.\ref{two} a simple derivation and analysis of the SCGA 
for the Ising systems without using diagrams is given.
In Sec.\ref{three} the classical spin diagram technique and construction 
of the SCGA for a general Hamiltonian (\ref{dham}) are described in more 
detail. In Sec.\ref{four} the analytic 
properties of the SCGA in different limiting cases are investigated, 
including the spherical limit, where the known results are generalized 
for the anisotropic Hamiltonian (\ref{dham}). In Sec.\ref{five} the results 
of the numerical solution of the SCGA equations for different classical spin 
models on different lattices are presented and compared with the available 
HTSE, LTSE, MC simulation, and $1/D$ expansion results, as well as with 
the experimental data on EuO and EuS. 
In Sec.\ref{six} some further applications of the SCGA 
and the possibilities of its generalization are discussed.

\section{IDEA OF the SCGA}
\label{two}

If in (\ref{dham}) the magnetic field ${\bf H}$ is directed along the 
ordering axis $z$ ($\eta_z=1$), then the $z$ component of the 
molecular field ${\bf H}_i$ acting on the spin on a site $i$ is given by
\begin{equation}\label{mfield}
H_{zi} = H + \sum_j J_{ij}m_{zj} .
\end{equation}
the MFA  consists in neglecting fluctuations of ${\bf H}_i$, which in the 
spatially homogeneous case leads to the Curie-Weiss equation for 
magnetization $m\equiv \langle m_z\rangle $:
\begin{equation}\label{cweiss}
m = B(\beta\langle H_z\rangle ), \qquad \langle H_z\rangle  = H + mJ_0 ,
\end{equation}
where $B(\xi)$ is the Langevin function, $\beta\equiv 1/T$, and $J_0$ is 
the zero Fourier component of the exchange interaction. The second 
moment of fluctuations of the $\alpha$ component of the molecular field 
${\bf H}_i$, which were neglected in the MFA , can be expressed as
\begin{eqnarray}\label{flumf}
\sigma_{2\alpha} = \sum_{jj'}J_{ij}J_{ij'}\eta_\alpha^2
\langle\Delta m_{\alpha j}\Delta m_{\alpha j'}\rangle  \nonumber \\
= v_0\!\!\!\int\!\!\!\frac{d{\bf q}}{(2\pi)^d}
(\eta_\alpha J_{\bf q})^2 S_{\alpha\alpha}({\bf q}) ,
\end{eqnarray}
where $\Delta {\bf m} \equiv {\bf m} -\! \langle m_z\rangle {\bf e}_z$,
$S_{\alpha\alpha}({\bf q})$ is the spin-spin 
cor\-re\-la\-tion function, 
$v_0$ is the unit cell volume, and $d$ is the lattice dimensionality. If 
correlations of spins on different lattice sites $j,j'$ are neglected, 
then for 
systems with nearest neighbor (equivalent neighbor) interactions the 
integral over the Brillouin zone in (\ref{flumf}) is proportional to 
$1/z$ and small for a large number of equivalent neighbors
$z$. This is justified in the temperature range 
$T\gg T_c$, but for $T\sim T_c$ the correlations in (\ref{flumf}) should 
be taken into account. For low-dimensional systems ($d=1,2$) the lattice 
integral in (\ref{flumf}) diverges with lowering temperature
at ${\bf q}=0$, which invalidates the MFA . 
For three-dimensional systems the magnitude of the 
molecular field fluctuations $\sigma_2$ remains finite and not very 
large, which is reflected by the shift of the actual values of $T_c$ in 
about 30\% downwards from $T_c^{\rm MFA}$ depending on the lattice structure 
and the details of spin interactions in (\ref{dham}). The latter 
makes feasible an improvement of the MFA  in $d\geqslant 3$ dimensions, which 
consists in taking into account molecular field fluctuations described 
{\em only} by the set of their second moments $\sigma_{2\alpha}$. This 
means that the averages of an arbitrary number of molecular field 
components decay pairwise, which is equivalent to the use of the {\em 
Gaussian} distribution function for the molecular field fluctuations. 
For the Ising model ($\eta_\alpha=0$ for $\alpha\neq z$)
this leads, in particular, to the expression for 
magnetization $m$ being given by a Langevin function with a spreaded 
argument:
\begin{eqnarray}%\label{bspr1}
m = \frac{1}{(2\pi\sigma_{2z})^{1/2}}\!\!\int\limits_{-\infty}^\infty\!\!
dH_{z,\rm fl}\exp
\left( -\frac{H_{z,\rm fl}^2}{2\sigma_{2z}}\right) \nonumber \\
\times B[\beta( \langle H_z\rangle  + H_{z,\rm fl} )]
\end{eqnarray}
or
\begin{equation}\label{bspr2}
m = \tilde B(\xi_z,l_z) = 
\frac{1}{\pi^{1/2}} \!\!\int\limits_{-\infty}^\infty \!\!dz\,
e^{-z^2} B(\xi_z + 2l_z^{1/2}z) ,
\end{equation}
where $\xi_z\equiv \beta(H+mJ_0)$ and $l_z\equiv\beta^2\sigma_{2z}/2$. 
To obtain a closed system of equations, one can calculate the 
spin-spin correlation function $S_{zz}({\bf q})$ in (\ref{flumf}) in the 
simplest Ornstein-Zernike approximation: 
\begin{equation}\label{spincorri}
S_{zz}({\bf q}) 
= \frac{\tilde B'(\xi_z,l_z)}
{1-\tilde B'(\xi_z,l_z)\beta J_{\bf q}} ,
\end{equation}
where the derivative of the Langevin function, $B'\equiv dB/d\xi$, 
is also renormalized by Gaussian fluctuations analogously to 
(\ref{bspr2}). This system of nonlinear 
equations for $m$ and $l_z$ given by (\ref{bspr2}), 
(\ref{spincorri}), and (\ref{flumf}) with $\alpha=z$ 
was obtained in a very technical manner by 
Horwitz and Callen \cite{horcal61} and was
solved numerically in Ref.\onlinecite{garlut84i}. 
Note that the integral over the Brillouin zone $\sigma_{2z}$, 
Eq. (\ref{flumf}), is taken into account 
in (\ref{bspr2}) in all orders of a perturbation theory. 
Such a self-consistent Gaussian approximation is, like all 
closed-form approximations in the theory of phase transitions, not a 
rigorous expansion in some small parameter. It is an approach taking 
into account some physically significant diagram structures 
self-consistently in all orders of a perturbation theory and reproducing 
the leading orders in the perturbatively treatable regions $T\ll 
T_c$ and $T\gg T_c$. In the next section the SCGA will be derived for a 
general form of the spin-vector Hamiltonian (\ref{dham}) with the use of 
the classical spin diagram technique.

\section{Classical spin diagram technique and the SCGA}
\label{three}

\begin{figure}%[p]
\unitlength1cm
\begin{picture}(11,6)
\centerline{\epsfig{file=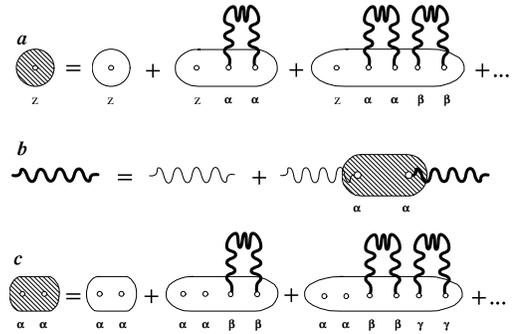,angle=0,width=12cm}}
\end{picture}
%\par
%
\caption{ \label{scgascga}
Self-consistent Gaussian approximation (SCGA) for classical spin 
systems.
(a) and (c): block summations for the renormalized magnetization 
and pair spin cumulant averages; 
(b): Dyson equation for the renormalized interaction line.
}
\end{figure}

This diagram technique can be considered as a simplified form of the 
quantum linked cluster expansion \cite{stietal63} or of the quantum SDT 
\cite{vlp6767,izykasskr74}, making use of the classical properties of spin 
vectors.
A perturbative expansion of the thermal average of any quantity ${\cal A}$ 
characterizing a classical spin system (e.g., ${\cal A} = m_z$) can be 
obtained by rewriting (\ref{dham}) as 
${\cal H} ={\cal H}_0 + {\cal H}_{\rm int}$, 
where ${\cal H}_0$ is the MFA  Hamiltonian with the molecular 
field $\langle H_z\rangle $ determined by (\ref{cweiss}), 
and expanding the expression 
\begin{equation}\label{statavr}
\langle{\cal A}\rangle  = \frac{1}{{\cal Z}} \int\prod_{j=1}^N d{\bf m}_j
{\cal A} \exp(-\beta {\cal H}), \qquad  |{\bf m}_j|=1 ,
\end{equation}
in powers of ${\cal H}_{\rm int}$. 
The integration in (\ref{statavr}) is carried out with respect to the 
orientations of the $D$-dimensional unit vectors ${\bf m}_j$ on each of the 
total $N$ lattice sites.
Averages of various spin vector components 
on various lattice sites with the Hamiltonian ${\cal H}_0$ can be 
expressed through spin cumulants, or semi-invariants, which will be 
considered below, in the following way:
\begin{eqnarray}\label{siteavr}
&& %\hspace{-1cm} 
\langle m_{\alpha i}\rangle _0  =  \Lambda_\alpha, \nonumber   \\
&&
\langle m_{\alpha i}m_{\beta j}\rangle _0 = \Lambda_{\alpha\beta} \delta_{ij} 
+ \Lambda_\alpha \Lambda_\beta,         \\
&& %\hspace{-1cm}
 \langle m_{\alpha i}m_{\beta j}m_{\gamma k}\rangle _0  =  
 \Lambda_{\alpha\beta\gamma} \delta_{ijk} 
+ \Lambda_{\alpha\beta}\Lambda_\gamma \delta_{ij} \nonumber  \\
&&  \qquad \qquad \qquad 
+\, \Lambda_{\beta\gamma}\Lambda_\alpha \delta_{jk}
+ \Lambda_{\gamma\alpha}\Lambda_\beta \delta_{ki}
+ \Lambda_\alpha \Lambda_\beta \Lambda_\gamma ,        \nonumber
\end{eqnarray}
etc., where $\delta_{ij}$, $\delta_{ijk}$, etc., are the site Kronecker 
symbols equal to 1 for all site indices coinciding with each other and to 
zero in all other cases. For the one-site averages 
($i=j=k=\ldots$) (\ref{siteavr}) reduces to the well-known 
representation of moments through semi-invariants, generalized for 
a multiple-component case. 
In the graphical language (see Fig. \ref{scgascga}) the decomposition 
(\ref{siteavr}) corresponds to all possible groupings of small circles 
(spin components) into oval blocks (cumulant averages). The circles 
coming from ${\cal H}_{\rm int}$ (the ``inner'' circles) 
are connected pairwise 
by the wavy interaction lines representing the quantity  
$\eta_\alpha\beta J_{ij}$ in (\ref{dham}). In diagram expressions 
summations over site indices 
$i$ and component indices $\alpha$ of inner circles are carried out. One 
should not take into account disconnected (unlinked) diagrams [i.e., those 
containing disconnected parts with no ``outer'' circles belonging to 
${\cal A}$ in (\ref{statavr})], since these diagrams are compensated for 
by the expansion of the partition function ${\cal Z}$ in the denominator 
of (\ref{statavr}). Consideration of combinatorial numbers shows 
that each diagram contains the factor $1/n_s$, where $n_s$ is the number 
of symmetry group elements of a diagram.   
The symmetry operations do not concern outer circles,
which serve as a distinguishable ``root'' to build up more complicated or 
renormalized diagrams. 
Such combinatorial factors are present, in particular, in the formulas  
(\ref{scga}) and (\ref{rencumlsmall}).
For practical calculations it is usually more 
convenient to use the Fourier representation and to calculate integrals 
over the Brillouin zone rather than lattice sums. As 
due to the Kronecker symbols in (\ref{siteavr}) lattice sums 
are subject to the constraint that the coordinates of the circles 
belonging to the same block coincide with each other, in the Fourier 
representation 
the sum of wave vectors coming to or going out of any block along 
interaction lines is zero. The cumulant spin averages in (\ref{siteavr}) 
can be obtained by differentiating the generating function $\Lambda(\xi)$ 
over appropriate components of the dimensionless field 
$\mbox{\boldmath$\xi$}\equiv \beta{\bf H}$ \cite{garlut84d}:
\begin{eqnarray}\label{defcum}
&&
\Lambda_{\alpha_1\alpha_2\ldots\alpha_p}(\mbox{\boldmath$\xi$})=
\frac{\partial ^p\Lambda(\xi)}{\partial \xi_{\alpha_1} 
\partial \xi_{\alpha_2}\ldots\partial \xi_{\alpha_p}}, \nonumber\\ 
&&
\Lambda(\xi)=\ln {\cal Z}_0(\xi),
\end{eqnarray}
where $\xi \equiv |\mbox{\boldmath$\xi$}|$,
\begin{equation}\label{2.7}
{\cal Z}_0(\xi)={\rm const}\times \xi^{-(D/2-1)}{\rm I}_{D/2-1}(\xi)
\end{equation}
is the partition function of a $D$-component classical spin, and 
${\rm I}_\nu(\xi)$ is the modified Bessel function. 
A similar technique was applied by L\"usher and Weisz \cite{luewei88npb} 
to generate HTSE's for a more general $n$-component $\phi^4$ model. 
For several lowest-order 
cumulants differentiation in (\ref{defcum}) leads to the following 
expressions:
\begin{eqnarray}\label{cum}
&&
\Lambda_\alpha(\mbox{\boldmath$\xi$}) = B_0(\xi)\,\xi_\alpha 
= B(\xi)\,\xi_\alpha/\xi ,                        \nonumber  \\
&&
\Lambda_{\alpha\beta}(\mbox{\boldmath$\xi$}) 
= B_0(\xi)\,\delta_{\alpha\beta} + B_1(\xi)\, \xi_\alpha \xi_\beta , \\
&&
\Lambda_{\alpha\beta\gamma}(\mbox{\boldmath$\xi$}) 
= B_1(\xi)\,( \xi_\alpha \delta_{\beta\gamma}
+           \xi_\beta  \delta_{\gamma\alpha}
+           \xi_\gamma \delta_{\alpha\beta} )      \nonumber \\
&& \qquad\qquad\qquad
+ \, B_2(\xi)\, \xi_\alpha \xi_\beta \xi_\gamma ,    \nonumber \\
&&
\Lambda_{\alpha\beta\gamma\delta}(\mbox{\boldmath$\xi$}) 
= B_1\, 3{\cal P}( \delta_{\alpha\beta}\delta_{\gamma\delta} ) \nonumber \\
&& \qquad\qquad\qquad
+ \, B_2\, 6{\cal P}( \xi_\alpha \xi_\beta \delta_{\gamma\delta} )
+ B_3\, \xi_\alpha \xi_\beta \xi_\gamma \xi_\delta ,    \nonumber
\end{eqnarray}
where $\delta_{\alpha\beta}$ is the spin component Kronecker symbol, 
${\cal P}$ is the symmetrization operator,
\begin{equation}\label{defbn}
B_n(\xi) \equiv 
\left( \frac{1}{\xi} \frac{\partial}{\partial \xi} \right)^n
\frac{B(\xi)}{\xi},
\end{equation}
and 
\begin{equation}\label{defb}
B(\xi) = d\Lambda(\xi)/d\xi = 
{\rm I}_{D/2}(\xi)/{\rm I}_{D/2-1}(\xi)
\end{equation}
is the Langevin function of $D$-component classical spins, 
which can be expressed through elementary functions 
for odd values of $D$: 
\begin{equation}\label{b135}
B(\xi) = \left\{
\begin{array}{ll}
\tanh(\xi),                             \qquad          & D=1 , 
\\
\coth(\xi) - 1/\xi,                     \qquad          & D=3 ,
\\
1/(\coth(\xi) - 1/\xi) - 3/\xi,         \qquad          & D=5 ,
\end{array}
\right.
\end{equation}
etc. The small- and large-argument expansions of the Langevin function 
$B(\xi)$ have the form
\begin{eqnarray}\label{bxismall}
&&
B(\xi) \cong \frac{\xi}{D} - \frac{\xi^3}{D^2(D+2)}
+ \frac{2\xi^5}{D^3(D+2)(D+4)}                   \\
&& \qquad
- \, \frac{5\xi^7}{D^4(D+2)(D+4)(D+6)}
\left( 1 + \frac{2}{5(D+2)} \right) + \ldots    \nonumber
\end{eqnarray}
and
\begin{equation}\label{bxigreat}
B(\xi) \cong 1 - \frac{D-1}{2\xi} + \frac{(D-1)(D-3)}{8\xi^2} + \ldots ,
\end{equation}
respectively. One can see from (\ref{bxismall}), that the functions 
$B_n(\xi)$, Eq. (\ref{defbn}), are all finite at $\xi=0$: 
$B_0(0)=1/D$, $B_1(0) = -2/[D^2(D+2)]$, $B_2 = 16/[D^3(D+2)(D+4)]$, etc. 
Accordingly, the spin cumulants $\Lambda_{\ldots}$ in (\ref{cum}) with an 
even number of coinsiding indices 
are given in this case by their first terms:
\begin{equation}\label{cumxizero}
\Lambda_{\alpha\alpha}=B_0(0), \qquad
\Lambda_{\alpha\alpha\beta\beta} = B_1(0)\,(1+2\delta_{\alpha\beta}), 
\end{equation}
etc., whereas all other cumulants turn to zero. At large arguments from 
(\ref{defbn}) and (\ref{bxigreat}) follows 
$B_n(\xi)\propto \xi^{-(1+2n)}$. In this limit all terms of 
(\ref{cum}) yield comparable contributions into $\Lambda_{\ldots}$, and a 
$k$-spin cumulant decays generally as 
$\Lambda_{\alpha_1\alpha_2\ldots\alpha_k} \propto \xi^{-(k-1)}$. If, however, 
the field $\mbox{\boldmath$\xi$}$ is directed along some axis $z$, then 
in the cumulant averages containing $z$ components of spins the leading 
terms can cancel each other. In particular, the two-spin cumulant 
$\Lambda_{\alpha\beta}$ in (\ref{cum}), which plays a big role in the 
following, can be rewritten explicitly as
\begin{equation}\label{cum2}
\Lambda_{\alpha\beta}(\mbox{\boldmath$\xi$}) 
= \frac{B(\xi)}{\xi}
\left( \delta_{\alpha\beta} - \frac{\xi_\alpha \xi_\beta}{\xi^2} \right)
+ B'(\xi) \frac{\xi_\alpha \xi_\beta}{\xi^2} .
\end{equation}
For $\mbox{\boldmath$\xi$}=\xi{\bf e}_z$ this expression simplifies to
$\Lambda_{zz} = B'(\xi)$ and $\Lambda_{\alpha\alpha} = B(\xi)/\xi$
($\alpha \ne z$). Now from (\ref{bxigreat}) one can see that,
for $\xi \gg 1$, $\Lambda_{zz} \propto \xi^{-2}$, whereas 
$\Lambda_{\alpha\alpha} \propto \xi^{-1}$.
 
The simplification of spin cumulants for 
$\mbox{\boldmath$\xi$}=\xi{\bf e}_z$ mentioned above takes place in the 
unrenormalized diagrams generated initially by the expansion of 
(\ref{statavr}) in powers of ${\cal H}_{\rm int}$ since 
there is only one nonzero component of the molecular field: 
$\xi_z = \xi = \beta (H + mJ_0)$. The complete 
form of spin cumulants (\ref{cum}), (\ref{cum2}) is needed, however, for 
the construction of the SCGA, which allows for both longitudinal and 
{\em transverse} fluctuations of the molecular field. The latter is the 
essense of the diagram technique for the multiple-component classical spin 
systems presented here.
In the Ising case the classical spin diagram technique coincides 
with the ``Ising part'' of the standard quantum SDT
\cite{vlp6767,izykasskr74,izyskr88} and can be used with Brillouin 
functions $B_S$ of a general spin $S$. In the 
Refs.\onlinecite{izykasskr74,izyskr88} 
the reader can find more technical details
concerning the construction of SDT for Ising systems, which play
the same role in the present classical SDT.  

Before proceeding to the construction of the SCGA we should make a remark 
about the numerical calculation of the generalized Langevin function 
$B(\xi)$ (\ref{defb}) for arbitrary $D$. One can see from (\ref{b135})
that for $D>1$ the function $B(\xi)$ contains terms divergent at 
$\xi\to 0$, although $B(\xi)$ itself is well behaved. This hampers 
numerical calculations, and the situation is aggravated for the 
derivative $B'(\xi)$ and for the functions $B_n(\xi)$ (\ref{defbn}) 
entering the spin cumulants (\ref{cum}), 
as well as for higher spin dimensionalities $D$.
The best way of calculating $B(\xi)$ is based on using 
the {\em backward} recursion relation with respect to $D$:
\begin{equation}\label{brecur}
B(D,\xi) = \frac{\xi}{D + \xi B(D+2,\xi)},
\end{equation}
which can be derived from (\ref{defb}) and the three-term recursion relation 
for the modified Bessel functions I$_\nu(\xi)$. This formula yields the 
proper small-argument behavior of $B(D,\xi)$, Eq. (\ref{bxismall}),
to leading order even for an inaccurate $B(D+2,\xi)$, and the proper 
behavior at $\xi \gg 1$ to leading order described by (\ref{bxigreat})
can be guaranteed, if we 
choose the first two terms of the large-$D$ expansion \cite{gar94jsp}
\begin{eqnarray}\label{bdgreat}
&&
B(\xi) \cong f(x) + \frac{1}{D}\frac{x}{1+x^2}f^2(x) 
+ O\left(\frac{1}{D^2}\right), \nonumber\\
&&
x\equiv 2\xi/D, \qquad f(x) \equiv \frac{x}{1+(1+x^2)^{1/2}} ,  
\end{eqnarray}
as an initial condition for the recurrence 
formula (\ref{brecur}). This procedure proves to be extremely good: 
Already one application of (\ref{brecur}) yields $B(\xi)$ with an
accuracy not worse than 0.6\% for $D=1$, 0.35\% for $D=2$, and 0.25\% for 
$D=3$ in the whole range of $\xi$,
and the process converges fast with the increase of the iterations number. 
\par
The self-consistent Gaussian approximation consists in taking into 
account {\em pair} correlations of the molecular field acting on a given 
spin from its neighbors, which implies the Gaussian statistics of 
molecular field fluctuations. The corresponding diagram sequence is 
represented in Fig. \ref{scgascga}
and is equivalent to the following closed system of nonlinear equations 
for magnetization 
$m\equiv \mbox{$\langle m_z\rangle $}$ and the normalized second moments 
$l_\alpha\equiv \beta^2\sigma_{2\alpha}/2!$ 
[cf. (\ref{flumf}) and (\ref{bspr2})] 
of the molecular field fluctuations:
\begin{eqnarray}\label{scga}
&&
m = \tilde\Lambda_z ,      \\
&&
l_\alpha = \frac{1}{2!}
v_0\!\!\!\int\!\!\!\frac{d{\bf q}}{(2\pi)^d}
\frac{\eta_\alpha \beta J_{\bf q}}
{1-\tilde\Lambda_{\alpha\alpha}\eta_\alpha \beta J_{\bf q}},
\qquad \alpha = 1,2,\ldots,D . \nonumber
\end{eqnarray}
Here the spin cumulants $\tilde\Lambda_{\ldots}$ renormalized by 
Gaussian fluctuations of the molecular field are given 
according to Figs. \ref{scgascga}(a), \ref{scgascga}(c) by the series

\begin{eqnarray}\label{rencumlsmall}
&&
\tilde\Lambda_{\ldots} = \Lambda_{\ldots}
+ \sum_{\alpha=1}^D \Lambda_{\ldots\alpha\alpha}l_\alpha  \\
&&\qquad
+\,\sum_{\alpha,\beta=1}^D 
\left(
1 - \delta_{\alpha\beta} + \frac{1}{2!}\delta_{\alpha\beta}
\right)
\Lambda_{\ldots\alpha\alpha\beta\beta} l_\alpha l_\beta
+ \ldots ,                                                   \nonumber
\end{eqnarray}
where the ``bare'' spin cumulants $\Lambda_{\ldots}$ are given by 
(\ref{defcum}) or (\ref{cum}). 
This series describing the influence of pair-correlated fluctuations of 
different components of the molecular field can be 
rewritten as
\begin{eqnarray}\label{rencum1}
&&
\tilde\Lambda_{\ldots}
= \prod_{\alpha=1}^D \sum_{n_\alpha=0}^\infty
\frac{1}{n_\alpha !}
\left(
l_\alpha\frac{\partial^2}{\partial \xi_\alpha^2}
\right)^{n_\alpha}
\Lambda_{\ldots}(\mbox{\boldmath$\xi$}) \nonumber\\
&&\qquad
= \exp
\left[
\sum_{\alpha=1}^D 
l_\alpha\frac{\partial^2}{\partial \xi_\alpha^2}
\right]
\Lambda_{\ldots}(\mbox{\boldmath$\xi$}) .
\end{eqnarray}
Such exponential differential operators were 
considered by Horwitz and Callen \cite{horcal61} for the Ising model. 
A generalization of their results for the multiple-component case yields
the closed formula
\begin{equation}\label{rencum2}
\tilde\Lambda_{\ldots} 
 = \frac{1}{\pi^{D/2}}\!\!\int\!\!d^{D} r \,e^{-r^2} 
\Lambda_{\ldots}(\mbox{\boldmath$\zeta$}),
\end{equation}
where $\mbox{\boldmath$\zeta$}$ is the spreaded molecular field given by
\begin{equation}\label{sprmf}
\mbox{\boldmath$\zeta$} \equiv \beta (H + mJ_0){\bf e}_z
+ 2\sum_{\alpha=1}^D l_\alpha^{1/2}r_\alpha {\bf e}_\alpha ,
\end{equation}
$
{\bf e}_\alpha$ is the unit vector in the direction $\alpha$, and the 
integration in (\ref{rencum2}) is performed with respect to the $D$-component 
vector variable ${\bf r} \equiv \{r_\alpha\}$. 
It can be seen that the renormalized spin cumulants
$\tilde\Lambda_{\ldots}$ are functions of $m$ and all $l_\alpha$.
In the Ising case the SCGA system of equations (\ref{scga}) reduces 
to the one obtained by Horwitz and Callen \cite {horcal61},
which was described in the preceding section, since here only 
$l_z$ in (\ref{scga}) is nonzero and $\tilde\Lambda_{zz}=\tilde B'(m,l_z)$.  
The expression for $l_\alpha$ in (\ref{scga}) differs from (\ref{flumf}) in 
that a zero term of the type $\int d{\bf q}J_{\bf q}\sim J_{ii}=0$ 
was added for convenience, which allows one to formulate the diagram
technique in terms of renormalized interactions. 

The number of unknown variables in the nonlinear system 
of the SCGA equations 
(\ref{scga}) is for a general form of the Hamiltonian (\ref{dham}) equal 
to $D+1$. Thus, for example, for a completely anisotropic Heisenberg model
there are four unknown variables: $m$, $l_x$, $l_y$ and $l_z$. 
In a more complicated case with the magnetic field transverse to 
the ordering axis $z$, which is not considered here, one should take 
into account different magnetization components and nondiagonal moments 
of molecular field fluctuations, $l_{\alpha\beta}$ with $\alpha\ne\beta$, 
and the number of unknowns increases considerably. Similar takes 
place for antiferromagnets in magnetic field and for multiple-sublattice 
structures. Of a practical interest is the case when in the Hamiltonian 
(\ref{dham}) there are groups of equivalent transverse ($\alpha\ne z$)
spin components having anisotropy factors $\eta_\alpha$, and hence the 
moments $l_\alpha$, equal to each other. In this case the number of 
unknowns in (\ref{scga}) diminishes; denoting such a group with the index 
$x$ and introducing 
$r_x^2 \equiv \sum_{\alpha\in x}r_\alpha^2$ and $n_x$ as the number of 
equivalent components in the group, one can simplify the 
$D$-dimensional integral in (\ref{rencum2}) with the help of the 
identity
\begin{equation}\label{rencumcyl}
\frac{1}{\pi^{n_x/2}}
\!\!\int\!\!d^{n_x} r_{\alpha\in x} e^{-r_x^2}\ldots 
= \frac{2}{\Gamma(n_x/2)}
\!\!\int\!\!dr_x r_x^{n_x-1} e^{-r_x^2}\ldots
\end{equation}
and make a replacement 
$\xi_\alpha^2 \Rightarrow \xi_x^2/n_x$, where 
$\xi_x \equiv 2l_x^{1/2}r_x$ and $l_x \equiv l_{\alpha\in x}$,
in the pair spin cumulants $\Lambda_{\alpha\alpha}$, 
Eq. (\ref{cum2}), entering (\ref{rencum2}). Thus, in 
particular, for the O($n$) model all components with $\alpha\neq z$ 
are equivalent, and there are three independent variables in the SCGA 
equations (\ref{scga}): $m$, $l_z$ and $l_x=l_\alpha$. The Gaussian integrals 
(\ref{rencum2}) reduce in this case to two-dimensional ones over 
$r_z$ and $r_x$, and in (\ref{rencumcyl}) $n_x=n-1$. Above 
$T_c$ in the absence of a magnetic field $m=0$ and all spin components are 
equivalent; there is only one unknown variable $l_z=l_\alpha$  in 
(\ref{scga}), and the intergal $\tilde\Lambda_{\alpha\alpha}$, 
Eq. (\ref{rencum2}), becomes one dimensional.

The SCGA system of equations (\ref{scga}) determines the equation of 
state of a magnetic system, i.e., the magnetization as a function of 
temperature and magnetic field, $m(T,H)$. The caloric properties of a 
magnetic system in the SCGA can be also determined. In particular, 
the energy of a spin system 
$U \equiv \mbox{$\langle{\cal H}\rangle $}$ can be
obtained by averaging the Hamiltonian (\ref{dham}) and using
the expression for the renormalized spin correlation function 
$S_{\alpha\alpha}({\bf q})$ determined by (\ref{flumf}) 
in the form
\begin{equation}\label{spincorr}
S_{\alpha\alpha}({\bf q}) 
= \frac{\tilde\Lambda_{\alpha\alpha}}
{1-\tilde\Lambda_{\alpha\alpha}\eta_\alpha \beta J_{\bf q}}
\end{equation}
[cf. (\ref{spincorri})].
Using the definition of $l_\alpha$ in (\ref{scga}) one gets
\begin{eqnarray}\label{defu}
&&
U = -Hm - \frac{1}{2}J_0 m^2
- \frac{1}{2} v_0\!\!\!\int\!\!\!\frac{d{\bf q}}{(2\pi)^d} J_{\bf q}
\sum_{\alpha=1}^D \eta_\alpha S_{\alpha\alpha}({\bf q}) \nonumber \\
&& \qquad
= -Hm - \frac{1}{2}J_0 m^2 
- T \sum_{\alpha=1}^D l_\alpha \tilde\Lambda_{\alpha\alpha} , 
\end{eqnarray}
i.e., the energy can be obtained as a by-product of the numerical solution
of the SCGA equations (\ref{scga}). Now the heat capacity 
$C_H=\partial U(T,H)/\partial T$ can be obtained by the differention 
of (\ref{defu}). The most strong result is, however, that for the free 
energy $F=-T\ln{\cal Z}$ of a system. Its diagrammatic derivation, which was 
accomplished by Horwitz and Callen \cite{horcal61} for 
the Ising model, is a rather complicated combinatorial problem,
since the free-energy diagrams have no distinguishable
outer circles, which could be used as a root for building 
renormalized diagrams. But the generalization of the corresponding 
results for the multiple-component case is straightforward and yields
\begin{equation}\label{deff}
\beta F = \frac{\beta}{2}J_0 m^2 - \tilde\Lambda
- \sum_{\alpha=1}^D L_\alpha 
+ \sum_{\alpha=1}^D l_\alpha \tilde\Lambda_{\alpha\alpha}, 
\end{equation}
where $\tilde\Lambda$ is the generating function of spin cumulants 
(\ref{defcum}) renormalized by Gaussian fluctuations 
[see (\ref{rencum2})] and
\begin{equation}\label{lbigalpha}
L_\alpha = -\frac{1}{2!}
v_0\!\!\!\int\!\!\!\frac{d{\bf q}}{(2\pi)^d}
\ln(1-\tilde\Lambda_{\alpha\alpha}\eta_\alpha \beta J_{\bf q}) .
\end{equation}
Considering in (\ref{deff}) $m$ and $l_\alpha$ as free parameters, i.e.,
$F=F(T,H,m,\{l_\alpha\})$, and using the identities 
$\partial\tilde\Lambda_{\ldots}/\partial\xi_\alpha = 
\tilde\Lambda_{\ldots\alpha}$ and  
$\partial\tilde\Lambda_{\ldots}/\partial l_\alpha = 
\tilde\Lambda_{\ldots\alpha\alpha}$, 
one can obtain the SCGA system of equations 
(\ref{scga}) from the requirement that $F$ be stationary with respect to 
$m$ and $l_\alpha$: $\partial F/\partial m =0$ 
and $\partial F/\partial l_\alpha =0$. The expression for the energy $U$,
Eq. (\ref{defu}), can be also obtained from (\ref{deff}): 
$U = \partial(\beta F)/\partial\beta$.

\section{ANALYTICAL PROPERTIES OF the SCGA AND THE SPHERICAL LIMIT}
\label{four}

In this section the behavior of the SCGA solution for classical spin 
systems is analyzed in the regions of high and low temperatures, in the 
spherical limit ($D\to \infty$) and in the vicinity of the critical 
point. It is convenient to choose the dimensionless temperature variable 
$\theta\equiv T/T_c^{\rm MFA}$, 
where $T_c^{\rm MFA}=J_0/D$, and the dimensionless 
magnetic field $h\equiv H/J_0$,  and susceptibility $\tilde\chi\equiv 
J_0\chi$. Then the (unspreaded) molecular field in (\ref{sprmf}) is 
written as $\xi_z = \beta (H+mJ_0) = (D/\theta)(h+m)$, and the quantities 
$l_\alpha$, Eq. (\ref{scga}), transform to
\begin{eqnarray}\label{lalpha}
&&
l_\alpha = \frac{D}{2\theta\tilde G_\alpha}
[P( \eta_\alpha\tilde G_\alpha ) - 1], \qquad
\tilde G_\alpha \equiv \frac{D}{\theta}\tilde\Lambda_{\alpha\alpha} ,
\nonumber \\ 
&&
P(X) \equiv 
v_0\!\!\!\int\!\!\!\frac{d{\bf q}}{(2\pi)^d}
\frac{1}{1-X\lambda_{\bf q}} ,
\end{eqnarray}
where $\lambda_{\bf q} \equiv J_{\bf q}/J_0$ 
satisfies $1\!-\!\lambda_{\bf q} \propto k^2$ for $a_0 k\ll 1$; $a_0$ is
the lattice spacing.  
The lattice integral $P(X)$ has the following properties:
\begin{equation}\label{plims}
P(X) \cong
\left\{
\begin{array}{lll}
1 + X^2/z,                         & X \ll 1 ,                 \\
W - c\,(\delta X)^{1/2},           & \delta X \ll 1,   & d=3 , \\
W - c\,\delta X \ln(c'/\delta X),  & \delta X \ll 1,   & d=4 ,       
\end{array}
\right. 
\end{equation}
where $\delta X \equiv 1-X$, $z$ is the number of equivalent neighbors 
and $W$ (the Watson integral) and $c,c'$ are lattice-dependent constants. 
For low-dimensional 
systems ($d=1,2$) the function $P(X)$ diverges for $X\to 1$; for
$d\geqslant 5$ 
the leading term of the expansion of $P(X)$ about $X=1$ is nonsingular. 
The values of the Watson integral $W$ are 1.34466 for the fcc lattice 
($z=12$), 1.39320 for the bcc lattice ($z=8$), 1.51639 for the sc 
lattice ($z=6$), 1.79288 for the diamond lattice ($z=4$), 1.23965 for the 
$d=4$ hypercubic (hpc) lattice ($z=8$), and 1.15631 for the $d=5$ hpc one 
($z=10$). For hypecubic lattices with $d\gg 1$ one has $W\cong 1+1/z$
with $z=2d$. 
The difference $W-1$ measures in the SCGA deviations from the 
molecular field behavior and tends to zero if $z\to \infty$.

In the high-temperature region ($\theta\gg 1$) the second moments of 
the molecular field fluctuations, $\sigma_{2\alpha}$, 
Eq. (\ref{flumf}), should 
be temperature-independent and, correspondingly, 
$l_\alpha\equiv \beta^2\sigma_{2\alpha}/2! \propto \theta^{-2}$.
In this case the renormalized 
spin cumulants $\tilde\Lambda_{\ldots}$ (\ref{rencum2}) are given for 
$l_\alpha\ll 1$ by the expansion (\ref{rencumlsmall}),
where the 1-st order terms written out correspond to diagrams 
with one integration loop 
in Figs. \ref{scgascga}(a), \ref{scgascga}(c). Now using 
(\ref{cumxizero}) one can calculate the quantity $\tilde G_\alpha$ 
in (\ref{lalpha}) in the lowest order: 
$\tilde G_\alpha \cong \theta^{-1} \ll 1$. Then from (\ref{lalpha}) and 
(\ref{plims}) follows 
$l_\alpha \cong \eta_\alpha^2 D/(2\theta^2 z) \ll 1$, which justifies 
the initial assumption. The latter can be used to find a more accurate 
value of $\tilde G_\alpha$ up to $\theta^{-3}$ from the first two terms 
of (\ref{rencumlsmall}) and (\ref{cumxizero}). 
This allows one to determine the reduced susceptibilities 
$\theta\tilde\chi_\alpha({\bf q}) = D S_{\alpha\alpha}({\bf q})$  
[see (\ref{spincorr})] in the SCGA up to $\theta^{-3}$. 
For a particular case of the \mbox{$n$-$D$} model 
($\eta_\alpha = 1$ for $\alpha \leqslant n$ and $\eta_\alpha = 0$ for
$\alpha > n$) we write down the {\em complete} expression for the longitudinal 
($\alpha \leqslant n$) reduced susceptibility up to $\theta^{-3}$, 
which can be obtained 
with the help of the classical SDT without using the SCGA and has the form
\begin{eqnarray}\label{chihtse}
&&
\theta\tilde\chi_\| \cong
1 + \frac{1}{\theta} + 
\left(1 - \frac{1}{z}\frac{n+2}{D+2} \right) \frac{1}{\theta^2} 
\nonumber \\
&& \qquad
+ \, \left(
1 - \frac{2}{z}\frac{n+2}{D+2} + \frac{2}{z^2}\frac{n+2}{(D+2)^2}
\right) \frac{1}{\theta^3} + \ldots                               
\end{eqnarray}
Here all terms except the last one are contained in the SCGA, the latter 
being relatively small as $1/[z(D+2)]$. Such a situation takes place in 
the high-temperature range for other thermodynamic functions [e.g., 
the energy (\ref{defu})], too, as well as in higher orders of a 
perturbation theory -- corrections to the SCGA are determined by two small 
parameters $1/z$ and $1/(D+2)$. It can be seen from (\ref{chihtse}) that
for the models with $n=D$ the dependence of $\chi_\|$ on
$D$ comes practically only from these correction terms and remains weak not
too close to $T_c$. In the SCGA the $D$ dependence of the reduced 
susceptibility of a spin system, as well as of its energy,  
appears only in the order $\theta^{-7}$ 
due to the last $1/(D+2)$-correction term in
(\ref{bxismall}) and is very weak. For this reason also the
values of $T_c$ determined in the SCGA from the divergence of susceptibility  
are for the models with $n=D$ very close to each other and to
the one of the spherical model. From the expression (\ref{chihtse}) it 
can be seen that in the case of a large number of spin components the 
susceptibility developes in a natural way in powers of $1/D$ and not of 
$1/n$, as was mentioned in the Introduction. The same is valid for other 
thermodynamic quantities as well.

In the low-temperature region ($\theta\ll 1$) the expansion of 
thermodynamic quantities in powers of $\theta$ is more complicated, 
because the longitudinal and transverse spin components are 
nonequivalent and all expressions depend on magnetization, which should 
be calculated self-consistently in each order. The small-fluctuation 
expansion (\ref{rencumlsmall}) is valid in the range $\theta \ll 1$, 
too, since the high-order spin cumulants diminish as appropriate powers 
of $1/\xi\propto\theta$ [see the discussion after (\ref{cumxizero})].
In the zero-field case, starting from $\xi=(D/\theta)\,m \cong D/\theta$, 
one can estimate different terms of the low-temperature expansion 
(\ref{rencumlsmall}) for the magnetization $m=\tilde\Lambda_z$. One gets
($\alpha\ne z$)
\begin{eqnarray}\label{cumlt}
&&
\Lambda_z = B \cong 1 - (D-1)/(2\xi) \cong 1 - \theta(D-1)/(2D), 
\nonumber \\
&&
\Lambda_{\alpha\alpha} = B/\xi \cong \theta/D, 
\nonumber \\ 
&&
\Lambda_{zz} = B' \cong (D-1)/(2\xi^2) \cong \theta^2(D-1)/(2D^2), 
\\
&&
\Lambda_{z\alpha\alpha} = (\partial/\partial\xi)(B/\xi) 
\cong - \theta^2/D^2, 
\nonumber\\  
&& 
\Lambda_{zzz} \cong \theta^3(D-1)/D^3 ,
\nonumber  
\end{eqnarray}
etc., the first of these formulas being the MFA expression for 
magnetization $m$ up to first order in $\theta$.
Now it can be seen that in the low-temperature range 
$\tilde G_z \cong (D/\theta)\Lambda_{zz} \propto \theta$, 
(\ref{lalpha}) yields $l_z \sim {\rm const}$, and the contribution of 
longitudinal fluctuations into $m$ given by 
$\Lambda_{zzz}l_z$ in (\ref{rencumlsmall}) is small as $\theta^3$. 
The leading contribution to $m$ comes from transverse fluctuations
(spin waves), since $\tilde G_\alpha \cong 1$ and 
$l_\alpha \cong [P(\eta_\alpha)-1] \,D/(2\theta)\gg 1$. 
For the absolute value of the second moment of the molecular field
fluctuations, $\sigma_{2\alpha}=2T^2l_\alpha$, Eq. (\ref{flumf}), 
the latter means $\sigma_{2\alpha} \propto \theta \ll 1$.
Now the  magnetization $m$ is given for $\theta \ll 1$ by the formula
\begin{equation}\label{mtheta}
m \cong 1 - \frac{\theta}{2D}\sum_{\alpha=2}^D P(\eta_\alpha)
\end{equation}
of the lowest-order spin-wave theory,
where the sum includes only transverse components. 
For the \mbox{$n$-$D$} model (\ref{mtheta}) simplifies to \cite{garlut84d}
\begin{equation}\label{mthetand}
m \cong 1 - \frac{\theta}{2D}
\left[(n-1)W +D-n \right]
\end{equation}
[see (\ref{plims})]. Such a linear dependence replaces 
for classical ferromagnets the quantum Bloch law 
$m \cong 1 - a\theta^{3/2}$.

In the next order of a perturbation theory in $\theta\ll 1$ 
with the help of (\ref{rencumlsmall}) and (\ref{mtheta}) one gets
\begin{equation}\label{gtilde}
\tilde G_\alpha \cong 1 - (\theta/D)[P(\eta_\alpha)-1] ,
\end{equation}
which in the case $\eta_\alpha=1$ leads for three-dimensional systems due to 
(\ref{plims}) to the singular negative contribution to $\l_\alpha$, 
Eq. (\ref{lalpha}), and, as a consequence, to a {\em positive} contribution to 
magnetization $\propto \theta^{3/2}$ 
in addition to the leading negative linear term in (\ref{mtheta}). 
The latter is an artifact of the SCGA 
related to the unbalanced renormalization of spin-spin correlation 
functions. This is, however, an effect of the next order of magnitude, 
which is suppressed by the magnetic field or in the anisotropic case 
$\eta_\alpha < 1$.

In the spherical limit $D\to\infty$, the SCGA becomes exact, since all other 
more complicated diagrams die out 
\cite {garlut84d,gar94jsp} as at least $1/D$. 
The Langevin function (\ref{defb}) simplifies in this limit to the first 
term of the formula (\ref{bdgreat}).
The expression for the square of the {\em spreaded} value of the
scaled argument $x=2\zeta/D$ in (\ref{rencum2}) reads
\begin{equation}\label{xspread}
x^2 = 
\left[\frac{2}{\theta}(h+m) + \frac{4}{D} l_z^{1/2} r_z \right]^2
+ \frac{16}{D^2}\sum_{\alpha=2}^D l_\alpha r_\alpha^2 .
\end{equation}
Since, according to (\ref{lalpha}), $l_{\alpha,z} \propto D$, 
the spreading of the $z$ component of the 
molecular field in (\ref{xspread}) can be neglected, whereas the 
transverse contributions to (\ref{xspread}), each of them is small as 
$1/D$, are essential due to their number of the order $D$. The 
renormalized cumulants $\tilde\Lambda_{\alpha\alpha}$ 
($\alpha\ne z$) entering the SCGA 
equations (\ref{scga}) are in the limit $D\to\infty$ all equal to each 
other and given according to (\ref{cum})
by $\tilde\Lambda_{\alpha\alpha} \cong \tilde B_0$, 
so that we can introduce $G = (D/\theta)\tilde B_0$.
The Gaussian integrals (\ref{rencum2}) are 
for $D\gg 1$ easily calculated by applying the identity
\begin{equation}\label{gauident}
\frac{1}{\pi^{1/2}} \!\!\int\limits_{-\infty}^\infty \!\!
dx\; e^{-x^2}\! f(ax^2) \cong f(a/2), \qquad  a\ll 1,
\end{equation}
for an arbitrary function $f$ successively $D-1$ times. 
Thus, the integration leads simply to the 
replacement $r_\alpha^2 \Rightarrow 1/2$ in (\ref{xspread}), 
and the SCGA equations (\ref{scga}) reduce after some transformations to
\begin{equation}\label{sphereq1}
\frac{m}{h} = \frac{G}{1-G}  
\end{equation}
and 
\begin{equation}\label{sphereq2}
1 - m^2 =  G \, \frac{\theta}{D} \sum_{\alpha=2}^D P(\eta_\alpha G).
\end{equation}
Comparing these results with (\ref{mtheta}) one can identify the
spherical model as a model which is described in the whole
temperature range by an effective lowest-order spin-wave theory. 
In a zero magnetic field ($h=0$) the magnetization $m$ 
disappears above $T_c$, and the 
quantity $G$, which can be determined from (\ref{sphereq2}), increases 
from 0 to 1 with lowering temperature from $\infty$ to $T_c$.
Below $T_c$ for $h=0$ from (\ref{sphereq1}) follows $G=1$, 
and $m^2$ determined 
from (\ref{sphereq2}) is a linear function of temperature. 
It turns to zero at the critical point
\begin{equation}\label{sphertc}
\theta_c^{(\infty)} 
= \left[ \frac{1}{D} \sum_{\alpha=2}^D P(\eta_\alpha) \right]^{-1},
\end{equation}
which reduces to the well-known result \cite{berkac52}
$\theta_c^{(\infty)} = 1/W$ in 
the isotropic case $\eta_\alpha = 1$ considered by Stanley 
\cite {sta68pr}. The corresponding result for the \mbox{$n$-$D$} model 
($n/D={\rm const}\mbox{$\leqslant 1$}$) was obtained 
in Ref.\onlinecite{garlut84d}.
The general formula (\ref{sphertc}), as well as the whole 
equation of state (\ref{sphereq1}), (\ref{sphereq2}), shows a crossover to 
the MFA behavior in the case $\eta_\alpha \to 0$, $\eta_z = 1$ , i.e., 
for the ``spherical Ising model''; see (\ref{dham}). In the
anisotropic case, i.e. for any $\eta_\alpha<1$, the singularity 
of the function $P(X)$ at $X=1$ [see (\ref{plims})] is suppressed, and 
the critical indices of the spherical model coincide with those of the MFA .  

Now we proceed to the investigation of the behavior of the SCGA solution for 
classical spin systems in the critical region. The first step is to 
search for $T_c$ as a point at which the longitudinal correlation function
given by (\ref{spincorr}) with $\alpha=z$ diverges at ${\bf q}=0$ for $h=0$. 
This leads to the condition 
$\tilde G_z\equiv (D/\theta)\tilde\Lambda_{zz} = 1$, 
which in the isotropic case 
($\eta_\alpha=1$) with the use of the symmetrization (\ref{rencumcyl})
can be transformed to the following nonlinear 
equation for $\theta_c$ \cite{garlut84d}:
\begin{eqnarray}\label{thetac}
&&
\theta_c 
= \frac{2D}{\Gamma(D/2)} 
\!\int\limits_{0}^\infty \!\!dr\;r^{D-1}\;{\rm e}^{-r^2}
F(2l_{c}^{1/2}r) ,    \nonumber    \\
&&
l_{c} = D(W-1)/(2\theta_c),
\end{eqnarray}
where  [cf. (\ref{cum2})]
\begin{equation}\label{fthetac}
F(\xi) = \left( 1-\frac{1}{D} \right) \frac{B(\xi)}{\xi} 
+ \frac{B'(\xi)}{D}.
\end{equation}
In the particular case $D=1$ this equation reduces to the one obtained by 
Horwitz and Callen \cite{horcal61} for the Ising model. As was 
stressed above by the analysis of the susceptibility HTSE, 
Eq. (\ref{chihtse}), 
this $\theta_c$ should be very close to that of the spherical model, the 
latter underestimating $T_c$ in (5-8)\% for three-dimensional systems. 
Equation (\ref{thetac}) can be solved analytically in two limiting cases:  
(i) for $D\gg 1$ using the $1/D$ expansion results of 
Ref.\onlinecite{gar94jsp} and (ii) for \mbox{$W-1 \ll 1$}, when 
the spreading of molecular field fluctuations in (\ref{thetac}) is 
small and the deviation from the spherical result is due to the 
last correction term in (\ref{bxismall}). In these limiting cases 
$\theta_c$ is given by
\begin{equation}\label{thetaclim}
\theta_c \cong \frac{1}{W}\times \left\{
\begin{array}{ll}
\displaystyle
\! 1 - \frac{2}{D}\,\frac{(W-1)^3}{W(2W-1)^2},    & \; D\gg 1 ,  \\
\displaystyle
\! 1 - \frac{2}{D+2}\,(W-1)^3,                    & \; W-1 \ll 1 .
\end{array}
\right.
\end{equation}
Considering the values of the Watson integrals $W$ 
listed after (\ref{plims}), one can see that, indeed, the
correction terms in (\ref{thetaclim}) are typically small.

An attempt to simplify the SCGA equations (\ref{scga}) about such a
defined transition temperature $\theta_c$ and to calculate the spontaneous 
magnetization $m$ just below $\theta_c$ shows that $\theta_c$ is 
actually the lower spinodal boundary of a fictitious first-order phase 
transition occurring in the SCGA due to its inaccuracy in a close critical 
region; i.e., the magnetization jumps to a finite value by crossing 
$\theta_c$ from above. This instability is due to the singular behavior 
of the function $P(X)$ near $X=1$ [see (\ref{plims})]: The decrease of 
$\tilde G_z$ from 1 below $\theta_c$ related to the increase of 
magnetization leads to a sharp decrease of molecular field fluctuations
and hence to a further increase of magnetization and so on. Analytically 
the absence of a continuous solution $m(\theta)$ below $\theta_c$
can be shown the most easily for the Ising model, where in the 
vicinity of $\theta_c$ the SCGA simplifies to a system of equations
\begin{eqnarray}\label{scgaaboutthetac}
&&
\delta \tilde G_z + (2D/\theta)\tilde B''' \,\delta l_z 
= 2[(D/\theta)\tilde B' -1],    \nonumber    \\
&&
\delta \tilde G_z + (1/(D+2))(D/\theta)^3 \tilde B'''\,m^2 = 0 .
\end{eqnarray}
Here the spreaded derivatives $\tilde B^{[n]}$  
are calculated with $l_{zc}=D(W-1)/(2\theta_c)$ and $\delta l_z$ is
determined as 
$\delta l_z = l_{zc} - 
D[P(1-\delta \tilde G_z) - 1]/[2\theta(1-\delta \tilde G_z)] > 0$. Below  
$\theta_c$ the right part of the first of Eqs. (\ref{scgaaboutthetac}) is 
positive, and this equation has no solution since the negative singular 
term with $\delta l_z$ ($\tilde B'''<0$) dominates over the positive one with 
$\delta \tilde G_z$. This is the case for lattice dimensions 
$d=3,4$; for $d\geqslant 5$ the situation depends on the numerical factors in 
(\ref{scgaaboutthetac}), and if (\ref{scgaaboutthetac}) has a solution, 
then the MFA behavior of the spontaneous magnetization 
with the critical index 
$\beta=1/2$ is reproduced. The latter is consistent with the
analysis of Larkin and Khmelnitski \cite{larkhm69}.

The breakdown of the SCGA in the close critical region shows that this 
approximation is more sensitive to critical effects than other 
closed-form approximations always reproducing the MFA behavior. In the next 
section it will be shown that the {\em upper} spinodal boundary of the 
phase transition determined from the temperature dependence of 
magnetization yields a much better approximation for $T_c$ than the lower 
one. It is so because the inverse susceptibility turns to zero 
at $T_c$ with  
zero derivative, and even small inaccuracies in determination of 
the susceptibility can exert a great effect on determination of $T_c$. 
On the contrary, 
inaccuracies in magnetization produce a smaller effect on $T_c$ due to 
the infinite slope of $m$  at $T_c$.
%\vspace{0.4cm}

\section{NUMERICAL RESULTS AND COMPARISONS}
\label{five}

The SCGA system of nonlinear equations (\ref{scga})  
was solved for different 
lattices and different types of the spin hamiltonan (\ref{dham}) by 
the Newton-Raphson iterative method. For the fcc, bcc,
sc and diamond lattices  the analytic expressions 
for the lattice integrals $P(X)$, Eq. (\ref{lalpha}), given by Joyce
\cite{joy717172} were used. For hypercubic
lattices $P(X)$ was reduced to one-dimensional integrals with
the modified Bessel function I$_0(x)$ and calculated
numerically. In other cases $P(X)$ was calculated by a direct
integration over the Brillouin zone ($k_{x,y,z}\in [0,\pi]$) with
the use of three-dimensional product quadratures composed of 
5- or 10-point one-dimensional Gaussian quadratures. The accuracy of these
quadratures is so high that one does not need to analytically
separate the divergence of the integrand in (\ref{lalpha}) at
$q\to 0$ for $X=1$.    
The Gaussian integrals (\ref{rencum2}) were calculated
for the Ising model with the use of the 5-, 6- or 8-point 
Gauss-Hermite quadratures, for the  \mbox{$x$-$y$}, plane rotator, 
and completely anisotropic Heisenberg models, with the
use of the corresponding product quadratures. 
For the models with equivalent spin
components, such as O($n$) with $n\geqslant 3$, the symmetrized
integrals of the type (\ref{rencumcyl}) were approximated by the
5-, 6- or 8-point generalized Gauss-Hermite quadratures 
corresponding to the weight function $|x|^\alpha\exp(-x^2)$ 
with $\alpha=1,2,3$ (Ref. \onlinecite{strsec66}) 
and $\alpha=2,4,6,8$ (Ref. \onlinecite{shachefra64}).
The latter was sufficient to calculate O($n$) models up to
$n=10$. The relative accuracy of calculations is not worse than 0.1\%,
which exceeds the intrinsic accuracy of the SCGA.        

\begin{figure}[t]
\unitlength1cm
\begin{picture}(11,6)
\centerline{\epsfig{file=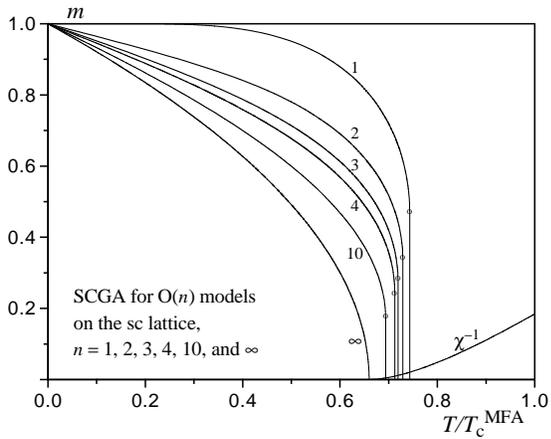,angle=-90,width=11cm}}
\end{picture}
%\par
%
\caption{ \label{scgamtdsc}
Temperature dependencies of spontaneous magnetization 
and zero-field susceptibility
of the O($n$) models on the sc lattice in the SCGA. 
}
\end{figure}

\begin{figure}[t]
\unitlength1cm
\begin{picture}(11,6)
\centerline{\epsfig{file=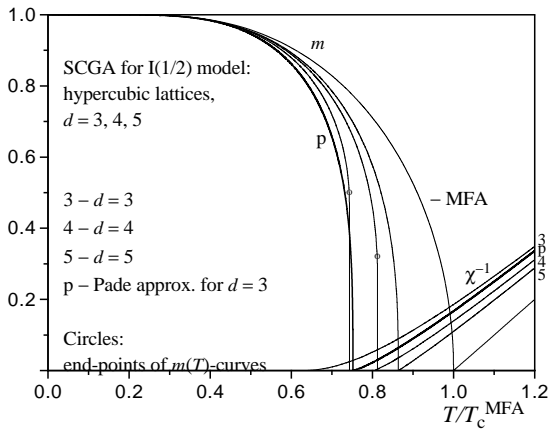,angle=-90,width=11cm}}
\end{picture}
%\par
%
\caption{ \label{scgamtihc}
Temperature dependences of spontaneous magnetization 
and zero-field susceptibility of the $S=1/2$ Ising model on
$d$-dimensional hypercubic lattices in the SCGA,  
compared with Pade-approximations of 
Refs.\protect\onlinecite{essfis63,bak61} for $d=3$.  
}
\end{figure}

The results represented in Table I, 
Fig. \ref{scgamtdsc} and Fig. \ref{scgamtihc}  show that the 
deviations from the MFA  due to molecular field fluctuations increase with 
the inverse of the number of interacting neighbors, $z$, or, 
rather, with the difference $W-1$ [see (\ref{plims})] 
depending on the lattice structure. Among the three-dimensional lattices
considered here the extreme cases are the diamond lattice ($z=4$) and
the equivalent neighbor fcc-sc lattice ($z=18$) with 12
face-centered-cubic nearest neighbors and 6 simple cubic next nearest ones.
For the O($n$) models ($n=D$)
the deviations of the magnetization $m$ from the MFA solutions
increase with the increase of $n$ (see Fig. \ref{scgamtdsc}): 
I(1/2) $\Rightarrow$ plane rotator $\Rightarrow$ H($\infty$) 
$\Rightarrow$ spherical model, whereas the susceptibilities 
$\tilde\chi$ are practically the same for all models. The latter
could be expected from the analysis of the susceptibility HTSE, 
Eq. (\ref{chihtse}), 
and of the lower spinodal boundary of the SCGA equations
(\ref{thetac}). For the \mbox{$n$-$D$} models with increasing $n$ and 
$D={\rm const}$ 
[I($\infty$) $\Rightarrow$ classical \mbox{$x$-$y$}  $\Rightarrow$ 
H($\infty$)]
the deviations from the MFA  are increasing stronger than
for O($n$) ones. This feature is in accordance with the functional
form of the susceptibility HTSE's, Eq. (\ref{chihtse}).
 
The temperature 
dependences of magnetization and other thermodynamic quantities 
calculated with increasing temperature are smooth functions of $T$ up 
to some ``upper spinodal boundary'' after which in a zero magnetic field 
magnetization jumps to zero. This feature results from the inaccuracy 
of the SCGA in a close critical region and was discussed in more detail 
at the end of Sec. \ref{four}. 
The discontinuities of thermodynamic functions 
in the SCGA diminish with the increase of the 
number of interacting neighbors, $z$, and the number of spin
components, $D$, as well as with the decrease of the number of
interacting spin components, $n$. For the quantities which are 
less singular at the critical point (e.g., the energy; see 
Ref. \onlinecite{garlut86jpf}) these discontinuities
are essentially weaker than the corresponding deviations from the MFA  
described by the SCGA. 
\par
\par
As was mentioned at the end of the preceding section, the SCGA 
upper spinodal boundary should provide a good estimate of phase 
transition temperatures of three-dimensional systems. Indeed, the 
corresponding 
values of $\theta_c\equiv T_c/T_c^{\rm MFA}$ listed in Table I differ from 
the ones obtained by HTSE and other accurate methods 
generally by (1-0.5)\%, which was never 
achieved by some other closed-form approximation. One can see that for 
the models with higher spin dimensionalities, $D$, the accordance with 
HTSE results is better 
than that for the most critical for the SCGA $S=1/2$ Ising model ($n=D=1$).
One can see from the Table I, that the SCGA yields for the O($n$)
models with $n<10$ more accurate results than the $1/n$ expansion
performed for the fcc, bcc and sc lattices by Okabe and Masutani
\cite{okamas78}, who calculated numerically the analytical
expressions (the double integrals over the Brillouin zone)
obtained by Abe and Hikami \cite{abe73,abehik7377}.

The models with higher lattice dimensionalities, $d$, 
are very important for testing 
the present approximation, which allows for Gaussian, i.e., non-interacting, 
fluctuations in the system. The interaction between fluctuations
dies out in the spherical 
limit $D\to\infty$, as well as for $d\geqslant 4$ \cite{larkhm69}. Thus, 
one can expect that the SCGA yields rather accurate results for  
$d\geqslant 4$, whereas the deviations from the MFA  described by the SCGA are 
still appreciable. Indeed, from Fig. \ref{scgamtihc} one can see that  
for the I(1/2) model on the $d=4$ hypercubic lattice the distance between 
upper and lower spinodal boundaries is very small. The latter is due to 
the fact that the singularity of the function $P(X)$ at $X\to 1$ [see 
(\ref{plims})] is only logarithmic. The discontinuity of magnetization 
at $T_c$ is for the I(1/2) model still substantial, but 
decreases quickly with the increase of the number of spin components $D$.
For $d\geqslant 5$ discontinuities of thermodynamic functions in the SCGA 
disappear. The value of $\theta_c$ for the I(1/2) model in $d=4$
dimensions (see Table I)
is over 2\% less then 
the $1/d$ expansion result of Fisher and Gaunt \cite{fisgau64}, and this 
discrepancy diminishes smoothly with the increase of $d$. This can be seen 
from the comparison with the $\theta_c$ values of 
Ref.\onlinecite{fisgau64} for 
$d=5,6$ and the recent high-accuracy results of Ref.\onlinecite{gofetal93} for 
$d=6,7$. For $d=6,7$ the SCGA yields the $\theta_c$ values 
0.894 [0.90227 (Ref. \onlinecite{fisgau64}), 
0.90290 (Ref. \onlinecite{gofetal93})] and 
0.913 [0.91922 (Ref. \onlinecite{gofetal93})], respectively. 
These values of 
$\theta_c$ are also very close to those for the spherical model (see 
Table I). 
With the increase of the spin dimensionality the accuracy of the SCGA
also increases.
In Table I the SCGA-results for $\theta_c$ for $d=4,5$ 
are compared with those of the general-$n$ $1/d$ expansion  
by Gerber and Fisher \cite{gerfis74} 
terminated by the term $d^{-5}$.
Unfortunately, the terminated $1/d$ expansion becomes less accurate 
for larger values of $n$ and does not reproduce, unlike the SCGA,
the exact results for the spherical model. 
It should be also noted that for the O($n$) models with
$d\geqslant 4$ the results for $\theta_c$ approach with the
increase of $n$ those for the spherical model much faster than
in three dimensions. This means that the coefficient in the $1/n$
expansion for $\theta_c$ 
in Refs.\onlinecite{abe73,abehik7377,okamas78} should be very small in
high dimensions.

The values of the energy of 3-dimensional spin systems 
on the upper spinodal boundary of the SCGA equations are also 
rather close to the series ones. The calculated values of the normalized 
energy $\tilde U(\theta_c)\equiv U(\theta_c)/U(0)$ of the $S=1/2$ 
Ising model are 0.276 (0.25), 0.298 (0.27),
0.365 (0.33) for the fcc, bcc and sc lattices, where the HTSE 
results are placed in brackets for   
comparison. For the classical Heisenberg model the normalized critical 
energies are given by 0.237 (0.245), 0.261 (0.265), 0.315 (0.325), 
respectively. For systems with higher lattice dimensionalities, $d$, the
energies are close to the ones for the spherical model,
$\tilde U(\theta_c) = 1-\theta_c$, especially for
systems with many spin components. The 
normalized critical energies in the SCGA of the 
I(1/2), H($\infty$), and spherical models, respectively, are
0.225, 0.198, 0.1933 for $d=4$ and 0.143, 0.135, 0.1352 for $d=5$. 
\begin{table*}[p]
\caption{The values of reduced Curie temperatures
$\theta_c\equiv T_c/T_c^{\rm MFA}$ calculated for different
classical spin models on different lattices from the upper
spinodal boundary of the SCGA equations. The results of
other methods are placed below for comparison.}
\renewcommand{\arraystretch}{2.2}
\begin{tabular}{*{9}{l}}        % {||l||*{8}{l|}|}
%\hline
Model $\backslash$ Lattice \qquad 
& \parbox[t]{12ex}{fcc \\ ($z\!=\!12$)} 
& \parbox[t]{12ex}{bcc \\ ($z\!=\!8$)}
& \parbox[t]{12ex}{sc \\ ($z\!=\!6$)}
& \parbox[t]{12ex}{Diamond \\ ($z\!=\!4$)}
& \parbox[t]{12ex}{fcc-sc \\ ($z\!=\!18$)}
& \parbox[t]{12ex}{bcc-sc \\ ($z\!=\!14$)}                
& \parbox[t]{12ex}{hpc(4$d$) \\ ($z\!=\!8$)}
& \parbox[t]{12ex}{hpc(5$d$) \\ ($z\!=\!10$)}
\\
\hline \hline
\parbox[t]{15ex}{O(1) \\(Ising, $S\!\!=\!\!1/2$) }
& \parbox[t]{12ex}{0.808 \\ 0.81617 \cite{fis63tab}
                         \\ 0.81620 \cite{camdyk75prb}
                         \\ 0.81628 \cite{sauworjas75} }        
& \parbox[t]{12ex}{0.785 \\ 0.79385 \cite{fis63tab}}        
& \parbox[t]{12ex}{0.744 \\ 0.75172 \cite{fis63tab}
                         \\ 0.75100 \cite{camdyk75prb} }         
& \parbox[t]{12ex}{0.673 \\ 0.6760  \cite{esssyk63}}
& \parbox[t]{12ex}{0.854 \\ 0.8609  \cite{dalwoo69}}           
& \parbox[t]{12ex}{0.824 \\ 0.8307  \cite{dalwoo69}}                
& \parbox[t]{12ex}{0.812 \\ 0.83401 \cite{fisgau64}}    
& \parbox[t]{12ex}{0.864 \\ 0.87694 \cite{fisgau64}}
\\
\parbox[t]{15ex}{O(2) \\(plane rotator)}
& \parbox[t]{12ex}{0.797 \\ 0.8033  \cite{fermoowor73} }               
& \parbox[t]{12ex}{0.773 \\ 0.7802  \cite{fermoowor73}
                         \\ 0.78019 \cite{butcom95} }
& \parbox[t]{12ex}{0.729 \\ 0.7343  \cite{fermoowor73}
                         \\ 0.73389 \cite{butcom95}
                         \\ 0.7332  \cite{tomyon95}    }
& \parbox[t]{12ex}{0.651}
& \parbox[t]{12ex}{0.847}
& \parbox[t]{12ex}{0.813}                                       
& \parbox[t]{12ex}{0.810 \\ 0.8319 \cite{gerfis74} }
& \parbox[t]{12ex}{0.864 \\ 0.8762 \cite{gerfis74} }
\\
\parbox[t]{15ex}{O(3) \\(Heisenberg, \\ $S\!=\!\infty$)}
& \parbox[t]{12ex}{0.790 \\ 0.794  \cite{rusbakwoo74} 
                         \\ 0.7943 \cite{adlholjan93} }   
& \parbox[t]{12ex}{0.766 \\ 0.771  \cite{rusbakwoo74}
                         \\ 0.7705 \cite{adlholjan93}
                         \\ 0.77032 \cite{butcom95} }    
& \parbox[t]{12ex}{0.719 \\ 0.723  \cite{rusbakwoo74}
                         \\ 0.7216 \cite{adlholjan93} 
                         \\ 0.72148 \cite{butcom95} }    
& \parbox[t]{12ex}{0.637}            
& \parbox[t]{12ex}{0.842 \\ 0.8505 \cite{rusbakwoo74}}   
& \parbox[t]{12ex}{0.807 \\ 0.8145 \cite{rusbakwoo74}}  
& \parbox[t]{12ex}{0.808 \\ 0.8292 \cite{gerfis74} }
& \parbox[t]{12ex}{0.864 \\ 0.8747 \cite{gerfis74} }
\\
O(4)
& \parbox[t]{12ex}{0.785}   
& \parbox[t]{12ex}{0.761 \\ 0.76306  \cite{butcom95} }    
& \parbox[t]{12ex}{0.712 \\ 0.71239  \cite{butcom95}
                         \\ 0.7123   \cite{kankay95} }    
& \parbox[t]{12ex}{0.628}            
& \parbox[t]{12ex}{0.839}   
& \parbox[t]{12ex}{0.803}  
& \parbox[t]{12ex}{0.808 \\ 0.8275 \cite{gerfis74}
                         \\ 0.8210 \cite{hasetal87,luewei88plb} }
& \parbox[t]{12ex}{0.864 \\ 0.8737 \cite{gerfis74} }
\\
O(5)
& \parbox[t]{12ex}{0.781 \\ 0.798 \cite{okamas78} }   
& \parbox[t]{12ex}{0.757 \\ 0.771 \cite{okamas78} }    
& \parbox[t]{12ex}{0.707 \\ 0.728 \cite{okamas78} }    
& \parbox[t]{12ex}{0.621}            
& \parbox[t]{12ex}{0.837}   
& \parbox[t]{12ex}{0.799}  
& \parbox[t]{12ex}{0.807 \\ 0.8262 \cite{gerfis74} }
& \parbox[t]{12ex}{0.864 \\ 0.8730 \cite{gerfis74} }
\\
O(6)
& \parbox[t]{12ex}{0.778 \\ 0.789 \cite{okamas78} }   
& \parbox[t]{12ex}{0.754 \\ 0.762 \cite{okamas78}  
                         \\ 0.75295  \cite{butcom95} }    
& \parbox[t]{12ex}{0.704 \\ 0.717 \cite{okamas78}  
                         \\ 0.70009  \cite{butcom95} }    
& \parbox[t]{12ex}{0.616}            
& \parbox[t]{12ex}{0.835}   
& \parbox[t]{12ex}{0.797}  
& \parbox[t]{12ex}{0.807 \\ 0.8253 \cite{gerfis74} }
& \parbox[t]{12ex}{0.864 \\ 0.8724 \cite{gerfis74} }
\\
O(8)
& \parbox[t]{12ex}{0.774 \\ 0.778 \cite{okamas78} }   
& \parbox[t]{12ex}{0.749 \\ 0.751 \cite{okamas78}  
                         \\ 0.74640  \cite{butcom95} }    
& \parbox[t]{12ex}{0.698 \\ 0.702 \cite{okamas78}  
                         \\ 0.69221  \cite{butcom95} }    
& \parbox[t]{12ex}{0.608}            
& \parbox[t]{12ex}{0.832}   
& \parbox[t]{12ex}{0.793}  
& \parbox[t]{12ex}{0.807 \\ 0.8240 \cite{gerfis74} }
& \parbox[t]{12ex}{0.864 \\ 0.8717 \cite{gerfis74} }
\\
O(10)
& \parbox[t]{12ex}{0.771 \\ 0.771 \cite{okamas78} }   
& \parbox[t]{12ex}{0.746 \\ 0.745 \cite{okamas78}  
                         \\ 0.74184  \cite{butcom95} }    
& \parbox[t]{12ex}{0.694 \\ 0.694 \cite{okamas78}  
                         \\ 0.68680  \cite{butcom95} }    
& \parbox[t]{12ex}{0.603}            
& \parbox[t]{12ex}{0.830}   
& \parbox[t]{12ex}{0.790}  
& \parbox[t]{12ex}{0.806 \\ 0.8240 \cite{gerfis74} }
& \parbox[t]{12ex}{0.864 \\ 0.8711 \cite{gerfis74} }
\\
\parbox[t]{15ex}{O($\infty$) \\(spherical)}
& 0.74368
& 0.71777
& 0.65946
& 0.55776
& 0.81397
& 0.76656                                                       
& \parbox[t]{12ex}{0.80668 \\ 0.8150 \cite{gerfis74} }
& \parbox[t]{12ex}{0.86482 \\ 0.8674 \cite{gerfis74} }
\\
Ising, $S\!=\!1$
%\parbox[t]{15ex}{Ising, $S\!=\!1$}
& \parbox[t]{12ex}{0.844 \\ 0.851 \cite{domsyk57}
                         \\ 0.85246 \cite{camdyk75prb} 
                         \\ 0.85264 \cite{sauworjas75} }
& \parbox[t]{12ex}{0.826} 
& \parbox[t]{12ex}{0.790 \\0.79893 \cite{camdyk75prb} }
& \parbox[t]{12ex}{0.727}
& \parbox[t]{12ex}{0.883}
& \parbox[t]{12ex}{0.857}                                       
& \parbox[t]{12ex}{0.853}
& \parbox[t]{12ex}{0.896}
\\
\parbox[t]{15ex}{Ising \\ ($n\!=\!1$, $D\!=\!2$)}
& \parbox[t]{12ex}{0.845}
& \parbox[t]{12ex}{0.827} 
& \parbox[t]{12ex}{0.792}
& \parbox[t]{12ex}{0.730}
& \parbox[t]{12ex}{0.883}
& \parbox[t]{12ex}{0.858}                                       
& \parbox[t]{12ex}{0.854}
& \parbox[t]{12ex}{0.896}
\\
\parbox[t]{15ex}{Ising, $S\!=\!\infty$ \\ ($n\!=\!1$, $D\!=\!3$)}
& \parbox[t]{12ex}{0.868 \\ 0.874 \cite{domsyk57}
                         \\ 0.87682 \cite{camdyk75prb}
                         \\ 0.87698 \cite{sauworjas75} }
& \parbox[t]{12ex}{0.853}
& \parbox[t]{12ex}{0.822 \\ 0.83195 \cite{camdyk75prb} }
& \parbox[t]{12ex}{0.767}
& \parbox[t]{12ex}{0.902}
& \parbox[t]{12ex}{0.879}                                       
& \parbox[t]{12ex}{0.880}
& \parbox[t]{12ex}{0.916}
\\
\parbox[t]{15ex}{\mbox{$x$-$y$}, $S\!=\!\infty$ \\ ($n\!=\!2$, $D\!=\!3$)}
& \parbox[t]{12ex}{0.828 \\ 0.8354 \cite{fermoowor73}}               
& \parbox[t]{12ex}{0.808 \\ 0.8156 \cite{fermoowor73}}               
& \parbox[t]{12ex}{0.768 \\ 0.7760 \cite{fermoowor73}}
& \parbox[t]{12ex}{0.699}
& \parbox[t]{12ex}{0.871}
& \parbox[t]{12ex}{0.842}                                       
& \parbox[t]{12ex}{0.843}
& \parbox[t]{12ex}{0.890}
\\
\end{tabular}
\end{table*}
\par
\begin{figure}%[p]
\unitlength1cm
\begin{picture}(11,6)
\centerline{\epsfig{file=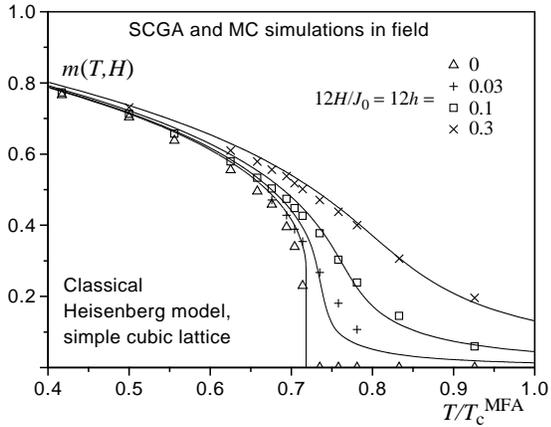,angle=-90,width=11cm}}
\end{picture}
%\par
%
\caption{ \label{scgamht}
Temperature dependences of magnetization in magnetic field $m(H,T)$ 
for the classical Heisenberg model in the SCGA, compared with 
MC-simulations of Ref.\protect\onlinecite{binmue73}. 
}
\end{figure}

In a magnetic field, $H\ne 0$, the SCGA becomes more accurate, because 
the system is driven away from the critical point $(0, T_c$), where the SCGA 
breaks down. The latter leads to the disappearance of the fictitious 
first-order phase transition in the SCGA starting from the fields, 
which are much 
smaller than the exchange interaction (i.e., for $h\equiv H/J_0 \ll 1$). For 
systems with a continuous spin symmetry (e.g., for the isotropic Heisenberg 
model) the magnetic field introduces a gap in the spin-wave spectrum and 
suppresses the singular contribution to magnetization 
$\propto \theta^{3/2}$ 
[see (\ref{gtilde}) and the following discussion], 
which improves the 
situation in the whole region below $T_c$. A comparison of the SCGA 
results for the magnetization in magnetic field $m(H,T)$ of the 
classical Heisenberg model on the sc lattice with the MC-simulation 
results of Binder and M\"uller-Krumbhaar \cite{binmue73} is represented 
in Fig. \ref{scgamht}.
\par
\par
\begin{figure}%[p]
\unitlength1cm
\begin{picture}(11,6)
\centerline{\epsfig{file=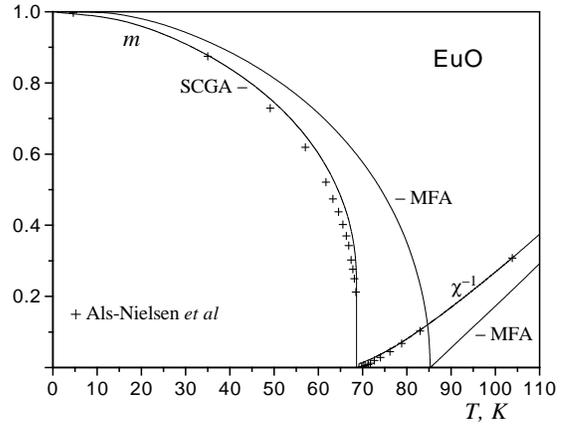,angle=-90,width=11cm}}
\end{picture}
%\par
%
\caption{ \label{scgaeuo}
Temperature dependences of magnetization and zero-field susceptibility  
of EuO in the SCGA, compared with the neutron scattering data 
of  Ref.\protect\onlinecite{alsdiepas76}.
}
\end{figure}
By application of the SCGA to experimentally investigated magnetic systems one 
should resrict 
oneself to the ones with large spin values ($S\gg 1$) and to the 
temperature range $T \gtrsim T_c/S$, where the whole Brillouin zone is 
populated by spin waves and the system behaves classically. An attempt 
to apply the SCGA to the $S\!=\!1/2$ Heisenberg model using the Brillouin 
function with $S=1/2$ [i.e., the Langevin function (\ref{defb}] with 
$D=1$), which corresponds formally to the consideration of the model with 
$n=3$ and $D=1$, yields for the sc lattice, in addition to the wrong 
linear behavior of the magnetization at low temperatures, 
the phase transition 
point $\theta_c=0.592$, being considerably higher than the HTSE 
value 0.560 \cite{rusbakwoo74}. On the other hand, for systems with 
$S\gg 1$ quantum effects in the range of elevated temperatures are 
determined by a small parameter \cite{vlp6767} $1/(zS)$  and can be 
partially taken into account in the SCGA by using the Brillouin function 
$B_S$. In typical cases this introduces errors that are smaller than 
the intrinsic inaccuracy of the SCGA. The Heisenberg ferromagnets EuO 
($T_c\simeq 69$ K) and EuS ($T_c\simeq 16.6$ K)
having $S=7/2$ are, perhaps, the most convenient materials for 
testing the SCGA, and they were extensively studied with NMR \cite{helben65} 
and neutron scattering \cite{alsdiepas76} methods. 
EuO and EuS form fcc lattices, and the exchange interaction extends 
up to the next nearest sc neighbors. The contribution of dipole-dipole 
interaction (DDI) to $T_c^{\rm MFA}$ is \cite{fisaha73} 1.7\% for EuO and 
4.9\% for EuS. With the use of HTSE it was shown \cite{jafwan79} that 
DDI suppreses to some extent the reduced transition temperatures 
$\theta_c\equiv T_c/T_c^{\rm MFA}$ due to its competing nature. 
In the SCGA rigorous taking into 
account DDI requires allowing for correlations between different 
components of molecular field fluctuations,
$l_{\alpha\beta}$ with $\alpha\neq\beta$,
which leads to the complication 
of the formalism and goes beyond the scope of this article. Instead of 
it, for a comparison with experimental data on EuO and EuS we
take DDI into account in a simplified manner, only through the 
renormalization of $T_c^{\rm MFA}$ mentioned above. The results of numerical 
calculations for EuO represented in Fig. \ref{scgaeuo} show the same 
accordance of the SCGA results with the neutron scattering data 
of Ref.\onlinecite{alsdiepas76} as its accordance  
with the HTSE and MC results demonstrated above. For EuS, due to the 
negative value of the n.n.n. exchange constant $J_2$, and hence the reduction 
of the effective number of interacting neighbours, the level of 
fluctuations is greater than in EuO, the Watson integral $W$ is close to 
that for the sc lattice, and the deviation of the results from the MFA , as 
well as the discrepancy between the SCGA and experimental results, is 
somewhat larger.

\section{Discussion}
\label{six}

The self-consistent Gaussian approximation (SCGA) for classical spin systems 
described here is a unified theory applicable to a wide class of
lattice models investigated currently by different groups with 
different methods. The SCGA takes into account fluctuations of the 
molecular field in the simplest way and is sensitive to the lattice 
dimensionality and structure and to the form of spin interactions.
The SCGA yields rather accurate values of 
the field-dependent magnetization, $m(H,T)$, 
and other thermodynamic functions 
in the whole plane $(H,T)$ excluding the vicinity of the critical point 
$(0,T_c)$. In particular, the values of $T_c$ themselves can be 
determined in the SCGA with an accuracy better than 1\%, which makes it 
already important for practical applications to new  lattice 
and spin Hamiltonian types. 

Indeed, the SCGA is more flexible (although less 
accurate) than series expansions, and consideration of new substances 
reduces in the simplest case to some modifications of the lattice 
integral $P(X)$, Eq. (\ref{lalpha}), and of the Gaussian integrals 
(\ref{rencum2}). 
This can be exemplified by studying the crossover between
fcc, sc, and bcc lattices varying the relative strength
of the first and second nearest neighbor interactions $J_2/J_1$,
or the crossover between Ising, Heisenberg, and \mbox{$x$-$y$} models
varying anisotropy constants. 
Having solved the SCGA system of equations (\ref{scga}), 
one obtains all quantities of interest 
as a result of a single calculation. More serious generalizations of 
the SCGA are required for consideration of systems with DDI or with a
transverse field, where nondiagonal correlations of molecular field 
fluctuations should be taken into account. For systems with 
many interacting sublattices the number of variables in the SCGA equations 
increases quadratically with the number of sublattices, and calculations 
become cumbersome.

Consideration of ferromagnets with a transverse field or 
antiferromagnets in field in the SCGA can be avoided, if one is interested
only in zero-field susceptibilities.
The zero-field 
ferro- and antiferromagnetic susceptibilities can be calculated through 
the correlation functions of the simplest ferromagnetic model with the 
longitudinal field. 
This requires, however, summation of some new diagram sequences
and is the subject of a separate work. 

Possible improvements of the SCGA should include 
non-Ornstein-Zernike effects in spin 
correlation functions (CF's) and non-Gaussian fluctuations of the molecular 
field. The former seems to be more important, since using 
Ornstein-Zernike CF's leads, due to singularities of the lattice integral 
$P(X)$, Eq. (\ref{plims}), to the overestimation of fluctuational effects for 
3-dimensional systems, which results in the breakdown of the SCGA 
in the critical 
region. The diagram technique for classical spin systems used for the 
construction of the SCGA is undoubtfully the best instrument for its further 
development, because it allows summation of different more complicated 
diagram series than those considered here. All other perturbative 
schemes that do not take explicitly the advantage 
of classical properties of a 
system fail to reproduce the SCGA, 
although the physical picture of Gaussian fluctuations of the molecular 
field is quite transparent. 

One more possible application of the classical spin diagram technique is 
that to low-dimensional and finite-size systems, where the level of 
fluctuations is large and an improvement of the SCGA is necessary. 
The first step in this 
direction was the calculation of the energy and susceptibility of 
low-dimensional antiferromagnets in the whole temperature interval 
\cite{gar94jsp} and also for a nonzero magnetic field 
\cite{gar95jmmm96jsp} with the use of the $1/D$ 
expansion. By this calculation, the results of which are rather good even for 
$D=3$, some diagram series going beyond the SCGA were summed up. This can, 
in principle, show how to improve the SCGA in a nonperturbative way
with respect to $D$.

The SCGA can be generalized also for inhomogeneous states of magnetics.
It turns out, however, that interesting results can be obtained already 
in the limit $D\to\infty$, where the model (\ref{dham}) is
analytically soluble but {\em not} equivalent to the standard
spherical model of Berlin and Kac \cite{berkac52} in  
inhomogeneous situations, even in the isotropic case. The
anisotropic spherical model defined by (\ref{dham}) in the limit
$D\to\infty$ was already applied to domain walls \cite{gar96jpa}
and to thin films \cite{gar96jpal}. 

And, finally, of a principal importance would be to construct the dynamical 
part of the classical spin diagram technique and to try to generalize 
the SCGA for dynamics.

\section*{ACKNOWLEDGMENTS}

The author thanks Hartwig Schmidt for valuable comments.
The financial support of Deutsche Forschungsgemeinschaft 
under contract No. Schm 398/5-1 is greatfully acknowledged.

%\bibliography{gar}

\end{document}